\documentclass[%
reprint,
amsmath,
amssymb,
aps,
prx,
]{revtex4-2}

\usepackage{xcolor}
\usepackage[colorlinks=true,citecolor=blue]{hyperref}
\usepackage{fancyhdr}
\usepackage{graphicx}
\usepackage{bookmark}
\usepackage{dsfont} 
\usepackage{upgreek}
\usepackage{booktabs}
\usepackage{tabularx}
\usepackage{listings}
\usepackage{multirow}
\usepackage{xspace}  
\usepackage[normalem]{ulem}  

\newcommand{\fsc}{\ensuremath{f_{\mathrm{SC}}}\xspace}
\newcommand{\frd}{\ensuremath{f_{\mathrm{RC}}}\xspace}
\newcommand{\frg}{\ensuremath{f_{\mathrm{RGC}}}\xspace}
\newcommand{\frlg}{\ensuremath{f_{\mathrm{RLGC}}}\xspace}
\newcommand{\fmg}{\ensuremath{f_{\mathrm{MGC}}}\xspace}
\newcommand{\fscinv}{\ensuremath{f^{-1}_{\mathrm{SC}}}\xspace}
\newcommand{\frdinv}{\ensuremath{f^{-1}_{\mathrm{RC}}}\xspace}
\newcommand{\frginv}{\ensuremath{f^{-1}_{\mathrm{RGC}}}\xspace}

\newcommand{\fmginv}{\ensuremath{f^{-1}_{\mathrm{MGC}}}\xspace}

\begin{document}
\begin{abstract}
    Generative quantum machine learning models are trained to deduce the probability distribution underlying a given dataset, and to produce new, synthetic samples from it. The majority of such models proposed in the literature, like the Quantum Circuit Born Machine (QCBM), fundamentally work on a binary level. Real-world data, however, is often numeric, requiring the models to translate between binary and continuous representations. We analyze how this transition influences the performance of quantum models and show that it requires the models to learn correlations that are solely an artifact of the way the data is encoded, and not related to the data itself. At the same time, structure of the original data, like continuity, can be obscured in the binary representation, hindering generalization. To mitigate these effects, we propose a strategy based on Gray-codes that can be implemented with essentially no overhead, conserves structures in the data, and avoids artificial correlations in situations in which the standard approach creates them. Considering datasets drawn from various low-dimensional probability distributions, we verify that, in most cases, QCBMs using the reflected Gray code learn faster and more accurately than those with standard binary code. This shows that, as complement to specifically tailored circuit ansätze or data pre-processing, binary encodings can be used to introduce inductive biases into generative machine learning models.
\end{abstract}
\pagestyle{fancy}
\title{Encoding Numerical Data for Generative Quantum Machine Learning}

\author{Michael Krebsbach}\email{michael.krebsbach@iaf.fraunhofer.de}
\affiliation{Fraunhofer Institute for Applied Solid State Physics IAF, Tullastraße 72, 79108 Freiburg, Germany}
\author{Florentin Reiter}
\affiliation{Fraunhofer Institute for Applied Solid State Physics IAF, Tullastraße 72, 79108 Freiburg, Germany}
\author{Thomas Wellens}
\affiliation{Fraunhofer Institute for Applied Solid State Physics IAF, Tullastraße 72, 79108 Freiburg, Germany}
\author{Hagen-Henrik Kowalski}
\affiliation{Bundesdruckerei GmbH, Kommandantenstraße 18, 10969 Berlin, Germany}
\author{Ali Abedi}
\affiliation{Bundesdruckerei GmbH, Kommandantenstraße 18, 10969 Berlin, Germany}

\maketitle

\section{Introduction} \label{sec:introduction}

Sampling from classically intractable probability distributions is a key benchmark demonstrating the computational potential of quantum computers \cite{bouland19on,arute19quantum,ransford25helios}. At the same time, the advent of generative machine learning (ML) shows that modeling probability distributions has numerous real-world applications. This naturally raises the question of whether generative quantum machine learning (QML) can help solve problems that are challenging for classical generative ML. Recent work provides a positive answer \cite{huang25generative}, highlighting the growing importance of developments in generative QML.

Popular generative QML algorithms include the quantum generative adversarial network (QGAN) \cite{lloyd18quantum,zoufal19} and the quantum circuit Born machine (QCBM) \cite{liu18,benedetti19a}. In the following, we will focus on the QCBM, although the results are equally applicable to other generative QML models. The QCBM encodes a probability distribution implicitly in the amplitudes of a quantum state generated by a parameterized quantum circuit, allowing efficient sampling through measurement.

There is a variety of implementations of QCBMs in the literature. Recent results demonstrate trainability for certain circuits up to thousand qubits and thousands of parameters \cite{recioarmengol25train}. Many of them focus on datasets that are specifically tailored to serve the needs of QCBMs, namely datasets consisting of bitstrings. This reduction to binary variables is often performed by dividing data points into \textit{below} or \textit{above} a certain threshold (e.g., for high-energy physics data \cite{rudolph24trainability} or images of handwritten digits \cite{recioarmengol25train}) or by directly designing tailored distributions like bars-and-stripes \cite{liu18,benedetti19a,he19,hamilton19generative,zhu19training,du20expressive,benedetti21variational,rudolph23synergistic,gili22evaluating,recioarmengol25train}, measurements from Ising-type experiments \cite{recioarmengol25train}, binary graph states 
\cite{bako24probleminformed,ballogimbernat25shallow}, or datasets of bitstrings with a specific Hamming weight \cite{coyle20,gili22do,rudolph23synergistic,rudolph24trainability}. However, classical real-world data is typically not binary. Images have gray-scales or color ranges, measurement results are often numerical and most datasets are represented as vectors of (floating point) numbers.

Dealing not only with binary data, but floating point numbers further complicates the training process, which can be swamped by barren plateaus or local minima in the first place \cite{rudolph24trainability,mcclean18barren,larocca25barren}. Not only does the model have to learn a non-trivial probability distribution, but it has to learn this distribution after the original data has been mapped to a binary space. 
Other works address this problem by modifying the circuit ansatz, for instance by selecting the entangling connections according to correlations in the binary data 
\cite{liu18,makarski25circuit,recioarmengol25train} or by preprocessing the data, such that it shows structures that can be easier to learn \cite{zhu22copula-based,zhu22generative,riofrio23, kiwit23}. 
We propose a complementary approach that can be implemented in addition to other preprocessing or circuit design strategies. This approach adjusts the mapping between numerical data and its binary representation. The choice of this map to bitstrings turns out to be crucial for the learning process. If chosen carelessly, it can obscure structure present in the original data and prevent generalization. Choosing this map carefully instead can conserve structure and present an important inductive bias to the model, enabling effective training without restricting the expressivity of the model. 

The term structure is used here to refer to any information about the data that is known or assumed in advance, like symmetry, monotonicity or specific correlations. The structure that we use as example in this work is that numerical data is often continuous and similar data points typically have similar probabilities. To exploit this structure, we turn toward so-called combinatorial Gray codes \cite{stiblitz43binary,gray53pulse,savage97a,mutze22combinatorial} which are lists of bitstrings in which neighboring bitstrings are similar to each other. As we will see below, the Gray property provides an inductive bias for continuous data, which is thus easier to learn for generative QML models. At the same time, this approach does not restrict the expressivity of models and still allows representing probability distributions on non-continuous data as well. 

This work is structured as follows: First, we introduce QCBMs, formally define binary codes and demonstrate that the standard binary code exhibits limitations which can be resolved using Gray codes in sec.~\ref{sec:methods}. We then proceed to show in simulations that this theoretical insight translates to improved QCBM training for various different probability distributions (i.e., centered Gaussian, multiple Gaussian and multiple saw-tooth, respectively), circuit ansätze, and loss functions in sec.~\ref{sec:results}. These results are discussed and concluded upon in sec.~\ref{sec:conclusion}.

\section{Methods} \label{sec:methods}

The task of generative (quantum) machine learning is to 
deduce the underlying probability distribution from a given dataset and to produce new samples from it. For this task, it is not important to explicitly construct the synthetic distribution but rather to be able to sample from it. As it turns out, parameterized quantum circuits are a natural fit for this task.

\subsection{Born Rule} \label{sec:born_rule}
The Born rule states that measuring an $n$-qubit quantum state $|\psi \rangle$ (in the $Z$-basis) returns the bitstring $b = b_{n-1} b_{n-2} \dots b_1 b_0 \in \{0, 1\}^n$ with probability
\begin{align}
    p(b) = |\langle b | \psi \rangle |^2. \label{eq:born_rule}
\end{align}
The state $|\psi \rangle$ therefore describes a probability distribution from which we can sample by measurement. Consequently, $|\psi \rangle$ can be written as
\begin{align}
    |\psi \rangle = \sum_{b\in \{0, 1\}^n} e^{i \varphi(b)} \sqrt{p(b)} |b\rangle, \label{eq:QCBMstate}
\end{align}
where the phases $\varphi(b)$ are additional degrees of freedom associated with each bitstring $b$ that cannot be observed by $Z$ measurements. The exact amplitudes of eq.\ \eqref{eq:QCBMstate} and, consequently, its probabilities $p(b)$ are typically unknown and obtaining them using state tomography requires an exponential number of measurements. Sampling a new bitstring $b$ from $p(b)$, on the other hand, requires only a single measurement.

The QCBM is an algorithm that uses eq.\ \eqref{eq:born_rule} to implicitly define a probability distribution 
\begin{align}
    p: \{0, 1\}^n \to [0, 1] ~. 
\end{align}
Its goal is to adjust the state $|\psi \rangle$ such that $p$ approximates some target distribution $q$ which is typically given by a dataset \cite{liu18,benedetti19a}
\begin{align}
    p \approx q ~. \label{eq:QCBM_target_1}
\end{align}

\subsection{Data representation}

The target probability distribution $q$ the QCBM is supposed to approximate can have many different forms. In particular the domain $\mathcal{D}$ of the dataset can be a variety of data spaces, ranging from binary data, to discrete or continuous numbers, graph states and categorical data. In particular, most data spaces are different from the binary space the synthetic distribution $p$ is defined on. In order to compare bitstrings $b \in \{0, 1\}^n$ sampled from $p$ and data points $x \in \mathcal{D}$ sampled from $q$ (potentially after a preprocessing step), the bitstrings $b$ have to be mapped to the data space $\mathcal{D}$ first.

Since there are $2^n$ possible measurement outcomes $b$ in eq.\ \eqref{eq:born_rule}, the QCBM can represent $2^n$ data points in total. Thus, the data space has to be discretized in some way (if it is not already discrete). Note that a one-hot encoding could only encode $n$ of the $2^n$ data points and therefore does not scale efficiently. Therefore, we choose $2^n$ points $x_0, \dots, x_{2^n-1} \in \mathcal{D}$ that we call \textit{representatives}. Each $x_j$ represents the portion of the data space $\mathcal{D}_j \subset \mathcal{D}$ that is more similar to it than to any other $x_k$, as measured by some positive semi-definite similarity measure $k(\cdot, \cdot)$ on $\mathcal{D}$ 
\begin{align}
    \mathcal{D}_j = \left\{ x \in \mathcal{D} \mid k(x, x_j) > k(x, x_k) \forall k \neq j \right\} ~. \label{eq:D_js}
\end{align}
As a convention, we assign data points that are equally similar to two or more representatives $x_j$ to the one with the largest index $j$. 

The standard way to proceed from here is to interpret a bitsting $b$ as index $j = \mathrm{int}(b)$ by the standard binary representation of integers and thereby associate it to the representative $x_{\mathrm{int}(b)}$ \cite{liu18,kondratyev21,makarski25circuit,gujju25llmguided,zhu22copula-based,zhu22generative,zhu19training,zhai22sampleefficientb,coyle21,du22,ganguly23implementing}. However, we generalize the QCBM at this point and define a \textit{binary code} to be a bijective function 
\begin{align}
    f: [0,1,2,\dots, 2^n-1] \to \{0, 1\}^n
\end{align}
that maps indices $j$ to their corresponding bitstrings $b = f(j)$. Assembling everything, $x_j$ is sampled from $p^f$ with probability 
\begin{align}
    p^f(x_j) = p(f(j)), \label{eq:binary_code_probability}
\end{align}
where the superscript $f$ indicates that the bitstring probability distribution $p$ from eq.\ \eqref{eq:born_rule} is mapped to a distribution on the discretized version of the original data space $\mathcal{D}$ by the binary code $f$.  

Similarly, we can define a discretized version of the target distribution
\begin{align}
    \hat{q}(x_j) = q(\mathcal{D}_j) = \int_{\mathcal{D}_j} {\rm d}x~q(x) \label{eq:q_hat}
\end{align}
and map it to $p$'s domain using $f^{-1}$:
\begin{align}
    q^{f^{-1}}(b) = \hat{q}(x_{f^{-1}(b)}) ~. \label{eq:q^f(b)_discretized}
\end{align}

Refining eq.\ \eqref{eq:QCBM_target_1}, we denote the quantum state created by the QCBM by $|\psi_\theta\rangle$, where $\theta$ indicates that it is a parameterized state. The goal of the QCBM is to train the parameters $\theta$ such that the model's implicit probability distribution $p_\theta$, mapped to $\mathcal{D}$ by the binary code $f$, approximates the discretized data distribution $\hat{q}$
\begin{align}
    p^f_\theta(x_j) \approx \hat{q}(x_j)
\end{align}
for all representatives $x_j$.

Note that the binary code $f$ is applied after any classical preprocessing of the data. Standard preprocessing steps such as normalization, principal component analysis, or copula transformations \cite{zhu22copula-based,zhu22generative,riofrio23, kiwit23} change the data distribution $q$ and the choice of representatives $x_j$, but they leave the map $f$ between indices and bitstrings untouched. The optimization of the binary code is therefore orthogonal to, and can be combined with, existing preprocessing pipelines. The same holds for circuit-design strategies that select entangling gates according to correlations in the binary data \cite{liu18,makarski25circuit,recioarmengol25train}. These adapt the circuit to the data that is encoded by a given $f$, whereas we ask which $f$ is best suited for structured datasets and given circuits. 

In the following section we discuss why it is not always recommendable to use the standard code and outline alternatives.

\subsection{Binary codes} \label{sec:binary_codes}
In this section, we will focus on one-dimensional numerical data $\mathcal{D} \subset \mathbb{R}$ and choose the representatives $x_j$ in an ordered way $x_0 < x_1 < \dots < x_{2^n-1}$. The multi-dimensional case is a vectorization of this one-dimensional case where each representative is a vector of such one-dimensional representatives $\vec{x}_{j_1,\dots, j_d} = (x_{j_1}, \dots, x_{j_d})^\top$.

\subsubsection{Standard Code}
As mentioned above, the intuitive choice for the binary code $f$ is binary counting, which defines the $i$th bit of the bitstring $b=b_{n-1} b_{n-2}\dots b_0$ to be
\begin{align}
    \fsc(j)_i = \left\lfloor \frac{j}{2^i} \right\rfloor \mod 2 ~,~ i=0,1,...,n-1 ~,
    \label{eq:standard_code}
\end{align}
where $\lfloor \cdot \rfloor$ denotes the floor function. We call this representation the standard code (SC). Its inverse is given by
\begin{align}
    \fscinv(b) = \sum_{i = 0}^{n - 1} b_i \cdot 2^i ~. \label{eq:standard_code_inv}
\end{align}
Descriptions of QCBMs typically choose this standard code $\fsc$ implicitly and directly associate the bitstring $b$ with its corresponding data point $x_{\fscinv(b)}$. But why should this particular code be the best choice out of $2^n!$ possible bijections between $[0, 2^n - 1]$ and $\{0, 1\}^n$? 

\subsubsection{Random Binary Code}
To illustrate the importance of this choice, consider a randomized representation, called a random code (RC) $\frd: [0, 2^n - 1] \to \{0, 1\}^n$. Since it is random and not created from a fixed set of rules, the full map would have to be saved, which would require an exponential amount of memory. More importantly, however, even if
the probability distribution $q$ on the data space $\mathcal{D}$ exhibits a certain structure, the corresponding distribution $q^{\frdinv}(b) = \hat{q}(x_{\frdinv(b)})$ 
of bitstrings would appear to be completely random. Obviously, it would be impossible to generalize to previously unseen data points, since these are mapped to random bitstrings, which are uncorrelated with the given dataset.

Returning to the standard code, we realize that it does not preserve \textit{closeness} between integers. While the integers 3 and 4 are neighboring integers, their corresponding bitstrings $\fsc(3) = 011$ and $ \fsc(4) = 100$ have a Hamming distance of 
\begin{align}
    H(011, 100) = 3    
\end{align}
which means that they are three bit flips apart. This means that representing both numbers with a quantum state, i.e. $|\psi\rangle=(|011\rangle+|100\rangle)/\sqrt{2}$,
requires entanglement between all three qubits. This small, local difference in the data hence requires a complicated, non-local representation in the state. On the other hand, other neighboring integers like $2$ and $3$ are mapped to neighboring bitstrings
\begin{align}
    H(010, 011) = 1 ~,
\end{align}
and representing them would not require an entangled state, but only a product state, i.e.,
$|\psi\rangle=|0\rangle |1\rangle(|0\rangle+|1\rangle)/\sqrt{2}$,
which can be generated by local operations.
Accordingly, the model has to learn correlations between bits that are not directly a consequence of correlations in the data, but rather of the way it is encoded. A shallow circuit ansatz could therefore have a bias toward distributions that assign similar probabilities to $x_2$ and $x_3$ but not to $x_4$. This imbalance increases for larger numbers of qubits, since the Hamming distance $H(\fsc(2^{n-1}-1), \fsc(2^{n-1}))=n$. On average, the Hamming distance between two bitstrings representing neighboring data points in the standard code quickly converges to 2 for large $n$. For details see Appendix \ref{sec:appendix:average_Hamming}. 

Target distributions that assign similar probabilities to similar data points, for example because they describe continuous processes, would benefit from binary codes $f$ that preserve \textit{closeness} in the data by mapping close data points to close bitstrings (with respect to the Hamming distance $H$). This would make such data easier to learn and to generalize.

\subsubsection{Gray Codes}
Similar problems have emerged at early stages of electronic data processing where increasing an integer by one, by flipping switches according to the binary representation, could lead to seemingly random outputs. If, for example, the switches were not perfectly synchronized, it could happen that data was read out when some bits were already flipped while others were still in their previous state. As a solution to this problem, a binary code was proposed that requires only a single switch to be flipped whenever an integer is increased by one \cite{gray53pulse}. Since then, a rich field of research has formed under the name of combinatorial Gray codes (see References \cite{savage97a,mutze22combinatorial,knuth21the} for reviews), with applications in data encoding \cite{ludman81gray}, compression \cite{richards86data} and storing \cite{chang92symbolic}, puzzles \cite{gros72theorie,gardner72mathematical} and error correction \cite{hammons94the}. More recently, Gray codes have been picked up in the field of quantum computing for the decomposition of quantum gates \cite{vartiainen04efficient,jones24decomposing}, variational quantum eigensolvers \cite{dimatteo21improving,siwach21quantum} and adiabatic quantum computing \cite{chang22improving}. 

Gray codes deal with lists of bitstrings, where adjacent bitstrings differ only by small amounts, and generalizations thereof. The defining feature of Gray codes $f_G$ is that they map neighboring integers to neighboring bitstrings
\begin{align}
    H(f_G(j), f_G(j + 1)) = 1 \label{eq:Gray_Code_Property}
\end{align}
for $0 \leq j < 2^n$. Here, the term neighboring refers to numerical distances between indices and Hamming distances between bitstrings, respectively. Viewing the space of bitstrings as $n$-dimensional hypercube, a Gray code is a path through the hypercube that moves along the edges of the cube and visits every corner exactly once, which is also called a Hamiltonian path \cite{gilbert58gray}. In the following, we list a selection of Gray codes and discuss their applicability for generative quantum machine learning.

\subsubsection{Reflected Gray Code} \label{sec:RGCC}
The most well-known Gray code is the reflected (binary) Gray code (RGC) \frg \cite{stiblitz43binary,gray53pulse}
\begin{align}
    \frg(j) = \fsc(j) \oplus R[\fsc(j)] ~, \label{eq:rgc}
\end{align}
where $\oplus$ is the bitwise xor operation and $R[\cdot ]$ the right-shift operation that cuts off the least-significant bit and pads a $0$ as new most-significant bit 
\begin{align}
    R[\fsc(j)] = \fsc(\lfloor j / 2 \rfloor) ~.    
\end{align}
\fsc and \frg coincide in the most-significant bit, which can be used to iteratively invert eq.\ \eqref{eq:rgc}, resulting in the inverse $\frginv$. While it is a well-established fact that \frg is a Gray code, we give a short proof that it satisfies eq.\ \eqref{eq:Gray_Code_Property} in Appendix \ref{sec:appendix:2n_proof} for those interested. \frg and \fsc for $n=3$ are listed in Table \ref{tab:n3_binary_codes}.

\begin{figure}
    \centering
    \includegraphics[width=\linewidth]{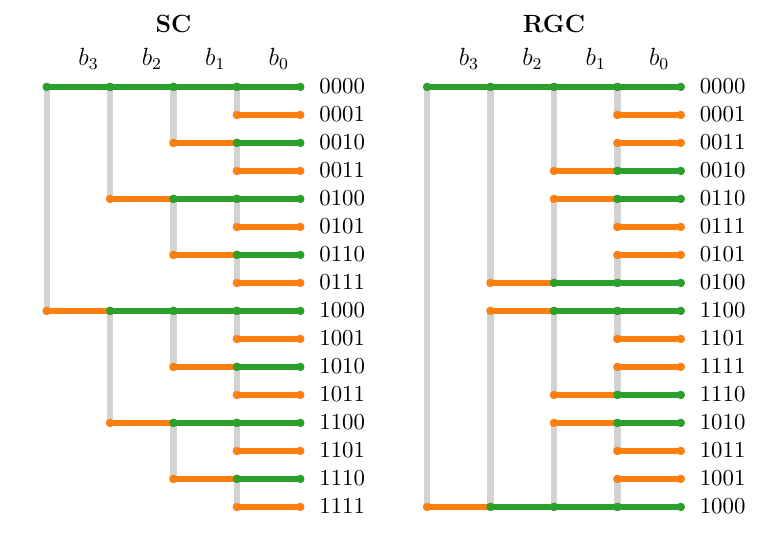}
    \caption{Binary tree representation of the standard (left) and binary reflected code (right) with $n=4$ bits. The bits are ordered from left ($b_3$, most significant) to right ($b_0$, least significant). A green line corresponds to $0$ and an orange line to $1$.}
    \label{fig:binary_tree}
\end{figure}

The term \textit{reflected} suggests that \frg exhibits reflection symmetries. Indeed, the $n+1$-bit code \frg can be constructed from the $n$-bit code \frg by writing it down as a list (e.g., $[0, 1]$ for $n=1$) and reflecting it at the end ($[0, 1, 1, 0]$). Adding a zero to the first half and and a one to the second, reflected half creates \frg for one more bit ($[00, 01, 11, 10]$). A visual representation of this is given in fig.~\ref{fig:binary_tree}, where \fsc and \frg are represented as binary trees for $n=4$ bits. This property suggests that \frg is especially well suited for data with mirror-symmetries, contrasting \fsc which exhibits translational symmetries.

\subsubsection{Maximum Run Length Gray Codes} 
The \textit{run length} of a Gray code is the maximal distance that is preserved by a code. It is defined as the maximal integer $r$ such that all distances smaller or equal $r$ are conserved:
\begin{align}
    H(\frlg(j), \frlg(j + k)) = k
\end{align}
for all $k \leq r$ and all $0 \leq j < 2^n-k$. \frg, for example, has a run length of $r=2$. While it is non-trivial to find binary codes of high run length, it is known that Gray codes with $r \geq n - 3 \log n$ exist \cite{goddyn03binary}. 

The quest for an isometric binary code mentioned above suggests that achieving a large run length $r$ is desirable, since this way distances are at least preserved locally. However, in the context of generative quantum machine learning on continuous data, this is a fallacy. A binary code with large run length $r \sim n$ enforces that similar integers $|j_1 - j_2| = r \ll 2^n$ are actually mapped to bitstrings with relatively large Hamming distance $r \sim n$. Superpositions of these bitstrings correspond to specific, highly entangled states that are typically non-trivial to learn. 

\subsubsection{Monotone Gray Codes}

Another variant of Gray codes are monotone Gray codes (MGCs) \cite{savage95monotoneb}. They attempt to increase the Hamming weight $h$ (the Hamming distance to the all-zero bitstring which is the number of ones in $\fmg(j)$) almost monotically with $j$. Note that strict monotonicity is excluded by the Gray property requiring the Hamming weight of bitstrings corresponding to neighboring integers to differ by $\pm 1$. This means that for all integers $j$ and $k$ with $0 \leq j \leq k < 2^n$ 
\begin{align}
    h(\fmg(k)) \geq h(\fmg(j)) - 1 ~. \label{eq:almost_monotonous}
\end{align}
Once it reaches a Hamming weight of $h(\fmg({j}))$, \fmg never produces bitstrings with less than $h(\fmg(j))-1$ ones.

MGCs that comply with eqs.\ \eqref{eq:almost_monotonous} and \eqref{eq:Gray_Code_Property} exist for all $n\geq 1$ \cite{savage95monotoneb}. Starting from the all-zero bitstring, they oscillate between Hamming weights $\tilde h$ and $\tilde h + 1$ until no bitstring of Hamming weight $\tilde h$ is left. Then they continue oscillating between $\tilde h+1$ and $\tilde h+2$, ending with the all-one bitstring if $n$ is odd, or with a bitstring of Hamming weight $n-1$ if $n$ is even. For $n \geq 5$, there are multiple MGCs. We use the implementation of \cite{cromieriijima10alternative}, based on the proof in \cite{savage95monotoneb} as \fmg.  

In the quantum setting, the monotone property translates representatives $x_j$ with low index $j$ to basis states of low excitation $|\fmg(j) \rangle$ and those with high indices to states with high excitation. 

Despite both fulfilling eq.\ \eqref{eq:Gray_Code_Property}, \frg and \fmg differ in various details. First of all, \fmg does not show \frg's symmetry and might therefore not be as suited for symmetric data. Furthermore, it lacks the hierarchy between bits that \frg and \fsc exhibit. Instead, it establishes this almost monotonic relation between the excitation of states and the data points represented by them, which encodes the numerical structure of the data in a different way. Table \ref{tab:n3_binary_codes} lists \fsc, \frg and \fmg for $n=3$. 

In Section \ref{sec:results}, we investigate how the different codes influence the performance of QCBMs on various datasets. Before getting there, we need to define the circuit ansatz and the loss function used for training. 

\begin{table}
    \centering
    \caption{The standard, reflected Gray and monotone Gray codes for $n=3$.}
    \label{tab:n3_binary_codes}
    \begin{tabular}{ccccccccc}
    \toprule
    $i$ & ~0 & ~1 & ~2 & ~3 &~ 4 & ~5 & ~6 & ~7 \\
    \midrule
    $\fsc(i)$ & ~000 & ~001 & ~010 & ~011 & ~100 & ~101 & ~110 & ~111 \\ 
    $\frg(i)$ & ~000 & ~001 & ~011 & ~010 & ~110 & ~111 & ~101 & ~100 \\
    $\fmg(i)$ & ~000 & ~001 & ~011 & ~010 & ~110 & ~100 & ~101 & ~111 \\  \bottomrule
    \end{tabular}
\end{table}

\subsection{Parameterized Quantum Circuits} \label{sec:pqc}

\begin{figure}
    \centering
    \includegraphics[width=0.5\linewidth]{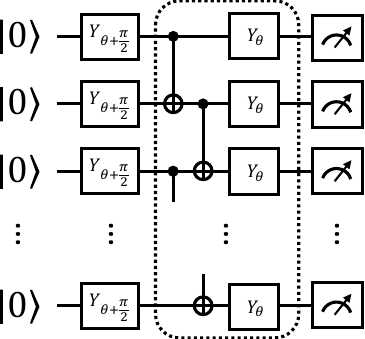}
    \caption{Schematic drawing of the circuit ansatz. It includes in total $n$ qubits initialized in the state $|0\rangle$ and the dashed box is repeated $L$ times. $Y_\theta$ represents a $R_Y$-rotation by the angle $\theta$ and each gate has its own parameter $\theta$. }
    \label{fig:circuit_ansatz}
\end{figure}

The circuit ansatz used in this paper is a straight-forward hardware-efficient ansatz (HEA) that is shown in fig. \ref{fig:circuit_ansatz}. It has $n$ qubits and $L + 1$ layers of parametrized single-qubit $R_y(\theta_i)$ rotations, interleaved with $L$ brickwork entangling layers that first apply CNOT gates between every even qubit $i$ and $i+1$, and then between $i+1$ and $i+2$. All gate parameters $\theta_i$, $1\leq i\leq n(L+1)$, are randomly initialized between $-0.025$ and $0.025$. The first rotation on each qubit gets an additional $R_y$ rotation by $\pi/2$ such that each qubit starts close to the state $|+\rangle = (|0\rangle + |1 \rangle)/\sqrt{2}$. Since the state $|{++}\rangle$ is invariant under CNOT operations and all single-qubit rotations are small, this initial configuration produces a distribution that is close to the uniform distribution of all bitstrings. With this linear topology, the light cone of each measurement extends to $\sim \min(4L, n)$ qubits for central measurements and to $\sim \min(2L, n)$ qubits for measurements at the end of the line.

In sec.~\ref{sec:Higherdim_data}, we additionally use an all-to-all circuit ansatz, which is similar to the HEA ansatz but replaces the nearest-neighbor entangling layers with all-to-all entangling layers that places CNOT gates between every pair of qubits $i$ and $j$ for all $i < j$. 

\subsection{Loss Function} \label{sec:loss_function}

For training a generative model,
a loss function that determines 
how well $p^f_\theta(x)$ approximates $q(x)$
is required. Typical (classical) loss functions like the Kullback-Leibler (KL) divergence
or the total variation (TV) distance
are so-called explicit losses and not well suited for training of implicit models, since they require knowledge of the complete probability distributions $q(x)$ and $p^f_\theta(x)$ \cite{rudolph24trainability}. Instead, the maximum mean discrepancy ($\mathrm{MMD}^2$) \cite{gretton08a,gretton12a} and Sinkhorn divergence ($\mathrm{SHD}$) \cite{feydy19, sinkorn64a, coyle20, coyle21} are typically used since they only depend on expectation values, which can be estimated from finite samples of these distributions. Note that the training data is drawn from $q(x)$ and not $\hat{q}(x)$. The squared MMD loss is defined as 
\begin{align}
    \mathrm{MMD}^2(p^f_\theta, q) = &\mathbb{E}_{x_j, x_k\sim p^f_\theta}[k(x_j, x_k)] \notag\\
    &+ \mathbb{E}_{x, y\sim q}[k(x, y)] \notag \\
    &- 2 \mathbb{E}_{x\sim q, x_j\sim p^f_\theta}[k(x, x_j)],
\end{align}
where $k(x, y)$ is the same similarity measure used to define $\mathcal{D}_j$ in eq.\ \eqref{eq:D_js}. We choose it to be the sum of $m$ Gaussian kernels
\begin{align}
    k(x, y) = \sum_{i=1}^m e^{-\frac{||x - y||_2^2}{2 \sigma_i^2}}, \label{eq:Gaussian_Kernel}
\end{align}
where $||x - y||_2$ is the two-norm between data samples and the $\sigma_i$'s are various bandwidth parameters. The choices of the bandwidth parameters $\sigma_i$ have a large influence on the trainability of QCBMs. For binary data, small bandwidths correspond to global observables, which can lead to barren plateaus, whereas large bandwidths correspond to local (or low-body) observables, for which shallow circuits are typically trainable \cite{rudolph24trainability}. For continuous data, the situation is slightly different. The $\mathrm{MMD}^2$ loss function compares the data on the numerical level. For a large bandwidth, it is a coarse-grain comparison of probability distributions, whereas for a small bandwidth, it has a high resolution. Since the data we use in this manuscript is bound between $-1$ and $+1$, large refers to $\sigma_i \in \mathcal{O}(1)$, and small to $\sigma_i \in \mathcal{O}(2^{-n})$. In order to relate it to the bodyness of the corresponding observables, similar to reference \cite{rudolph24trainability}, the numerical similarity has to be mapped to the binary level by the binary code. The bodyness of the observable, and therefore the trainability of the QCBM, heavily depends on the chosen binary code, preventing a general statement. However, heuristic considerations can still aid in choosing the bandwidth parameters. At initial stages of training, the model $p^f_\theta$ is likely very different from $q$. A high-resolution comparison with $\sigma_i \in \mathcal{O}(2^{-n})$ will therefore not detect any similarities, leading to a flat loss landscape. At initial stages of training, QCBMs therefore require large bandwidth parameters that compare the distributions on a high level. Once the model approaches the target distribution, the smaller bandwidths also detect similarities and lead to non-zero signals. These can be used for fine-tuning the model. Capable $\mathrm{MMD}^2$ functions should thus contain a combination of large, intermediate, and small $\sigma_i$, with respect to the relevant scales in the data. Following this guideline, we choose $m=5$ and $\sigma_i \in [0.003, 0.01, 0.03, 0.1, 0.3]$. Additionally, the above analysis suggests that the Gray property is particularly relevant when using small bandwidths, since it concerns local changes in the samples and the corresponding bitstrings. 

The loss function is optimized by the Adam optimizer \cite{kingma14adam} with parameters $\beta_1 = 0.9$ and $\beta_2 = 0.999$, and gradients are computed using the parameter-shift rule \cite{mitarai18quantum,schuld19evaluating}. Denote by $\theta_i^\pm$ the set of all parameters where parameter $\theta_i$ is shifted by $\pm \pi / 2$. The derivative of $\mathrm{MMD}^2$ with respect to $\theta_i$ is given by \cite{liu18}
\begin{align}
    \frac{\partial \mathrm{MMD}^2(q,p_\theta)}{\partial \theta_i} = &\mathbb{E}_{x_j \sim p_\theta^f, x_k \sim p_{\theta_i^+}^f}[k(x_j, x_k)] \notag \\
    &- \mathbb{E}_{x_j \sim p_\theta^f, x_k \sim p_{\theta_i^-}^f}[k(x_j, x_k)] \notag \\
    &- \mathbb{E}_{x \sim q, x_j \sim p_{\theta_i^+}^f}[k(x, x_j)] \notag \\
    &+ \mathbb{E}_{x \sim q, x_j \sim p_{\theta_i^-}^f}[k(x, x_j)] ~.
\end{align}
To compute the whole gradient, a total of $2 (L + 1) n + 1$ different circuits have to be evaluated. For each circuit 256 shots are used. This means that models with more qubits $n$ or deeper circuits $L$ require more resources for each gradient update step. 

The definition of the Sinkhorn divergence $\mathrm{SHD}$ and its gradient can be found in appendix \ref{sec:appendix:SHD}.

\section{Results} \label{sec:results}

\begin{figure*}
    \centering
    \includegraphics[width=\linewidth]{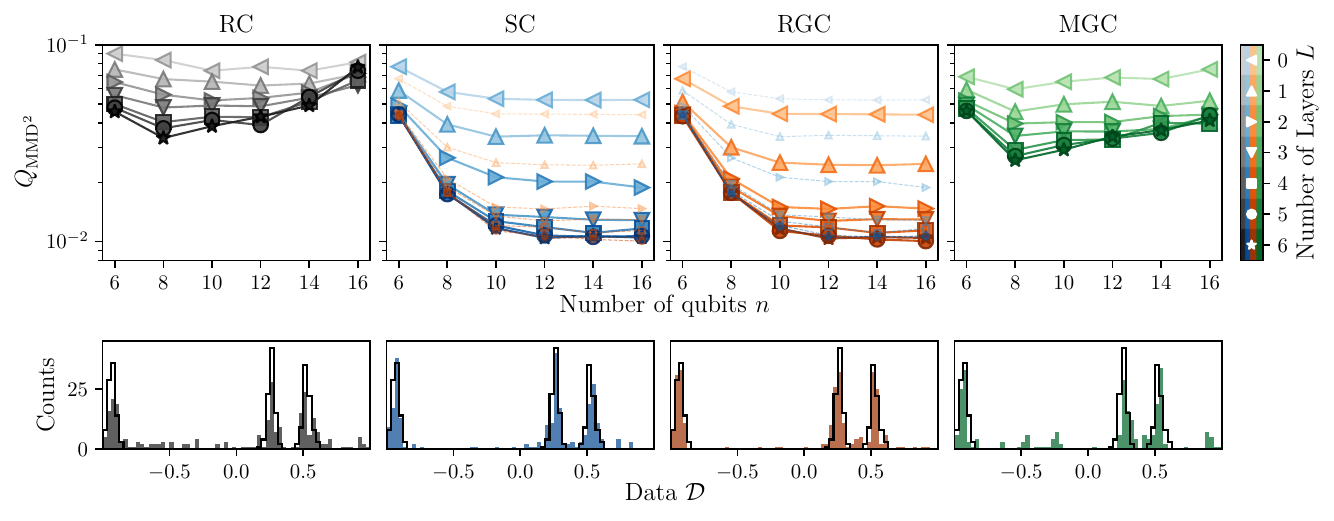}
    \caption{Various QCBMs trained on datasets drawn from three randomly placed Gaussian distributions with standard deviation $\nu = 0.03$. \textit{Top}: The quality score $Q_{\mathrm{MMD}^2}$, see eq.\ \eqref{eq:quality}, computed for the first 100 epochs for various models with different number of qubits $n=6, \dots, 16$ (x-axis), different number of layers $L=0, \dots, 6$ (marker and lightness, see legend on the right-hand side) and the four binary codes \frd, \fsc, \frg and \fmg (from left to right). Each point is the mean of 10 models trained on independently drawn datasets. For an easier comparison, the markers for \fsc are also drawn into the \frg plot and vice-versa as the small markers with dashed connection. 
    \textit{Bottom}: 256 training (black line) and 256 synthetic data points (bars) for \textit{one} model with $n=12$ qubits and $L=6$ layers trained for 100 epochs. For visibility, the data space $\mathcal{D}$ is discretized into $2^6$ (and not $2^{12}$) bins. }
    \label{fig:3_Blobs}
\end{figure*}

In this section, we investigate the effect of binary codes on the training of QCBMs. To that end, we train models on various different target distribution that are shown in figs.~\ref{fig:Symmetric_Blob}, \ref{fig:3_Blobs}, and \ref{fig:3_Sawtooth}, as well as of two- and three-dimensional generalizations thereof. 

\subsection{Centered Gaussian Distribution} \label{sec:Centered_Gaussian}
To illustrate the impact binary codes can have, we start the investigation with a very simple dataset consisting of 256 samples drawn from a single, one-dimensional Gaussian distribution of standard deviation $\nu=0.03$ that is centered around zero. The data space is chosen to be the interval $\mathcal{D} = [-1, 1] \subset \mathbb{R}$ and samples that lie outside this interval are set to $-1$ or $1$, respectively. 

\begin{figure*}
    \centering
    \includegraphics[width=\linewidth]{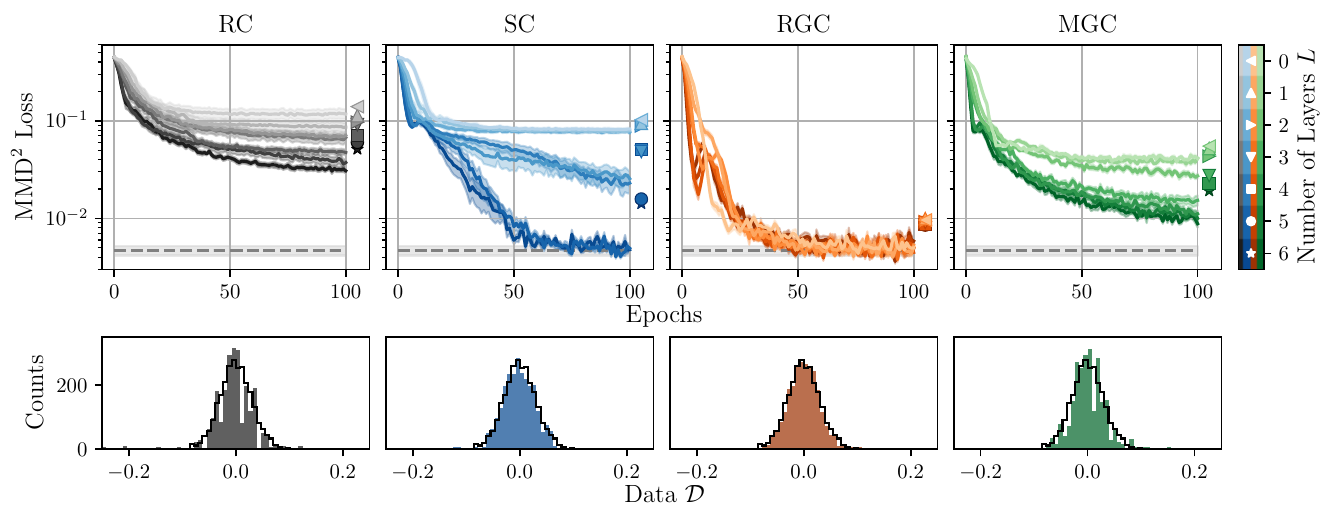}
    \caption{Training curves of various QCBMs with $n=8$ qubits and from left to right the random (gray), standard (blue), reflected Gray (orange) and monotone Gray code (green) trained on 256 samples drawn from a single Gaussian distribution with mean 0 and standard deviation $\nu = 0.03$. \textit{Top}: History of the $\mathrm{MMD}^2$ loss function during training. Each line represents the average of ten independent runs and the colored interval the standard error of the mean. For each binary code, circuits of various depths $L = 0, \dots, 6$ have been trained (lightness of the line). The markers at the end of each plot show the geometric mean of the loss $Q_{\mathrm{MMD}^2}$ over all epochs, indicating the model quality as it combines the convergence speed and achieved minimum in a single number. The gray dashed lines indicate the targeted reference value calculated by sampling and discretizing 256 additional samples from the target distribution, and comparing them to the training dataset. 
    \textit{Bottom}: The training (black line) and synthetic data (bars) of the models with $L=6$ after 100 epochs. The histograms show the aggregate of the ten independent runs, both in the training and the synthetic data, resulting in a total of $256 \times 10 = 2560$ samples. This aggregation indicates which patterns are exhibited consistently across all ten independent runs. Note that the $x$-axis is zoomed in on the relevant part of the data space. Each bin corresponds to one of the $2^8$ representatives $x_j$ and shows precisely $\mathcal{D}_j$. }
    \label{fig:Symmetric_Blob}
\end{figure*}

Fig.~\ref{fig:Symmetric_Blob} shows the training loss curves of various QCBMs with $n=8$ qubits in the top panels, as well as the synthetic data generated by them in the bottom panels. For each of the binary codes $\frd$, $\fsc$, $\frg$ and $\fmg$, QCBMs with between $L=0$ (no entanglement) and $L=6$ layers have been trained on ten independently sampled training datasets. The presented lines and colored areas represent the mean and standard error of the ten independent runs. Additionally, 256 test data points were drawn from the Gaussian and discretized into a distribution of the $2^8$ representatives $x_j$. Since they originate from the target probability distribution, they can be used to calculate the targeted $\mathrm{MMD}^2$ loss function value to be achieved by the generative model. While lower values are possible, they correspond to overfitting
(i.e., reproduction of the given dataset instead of the target distribution). Fig.~\ref{fig:Symmetric_Blob} shows the mean and standard error of this optimal value for the ten independently sampled datasets. 

The data shows that already for such a simple dataset, \frd does not achieve a similar accuracy as the other binary codes. It would require deep circuits with a large enough expressivity to make up for the missing structure in the random code. And even if a very deep circuit was able to recreate the training data, it would not be able to generalize to previously unseen data points, as discussed above.

The QCBMs with \fsc show a clear dependence on the number of layers $L$. The models with $L=0$, $1$ or $2$ entangling layers quickly get stuck at high loss function values. They lack the expressivity to represent the target probability distribution $q^{\fscinv}$ mapped to the binary space by \fscinv. Since the dataset is centered, \fsc maps the two most common central data points to two bitstrings $01111111$ and $10000000$ with maximal Hamming distance. Because the light-cones of the measurements of the most- and least-significant bits intersect only for $L\geq 3$, at least three layers are required to create synthetic data with correlation between all bits. The models with $L=3$ or $4$ layers indeed find a lower loss function value, however, they converge comparably slow. After 100 epochs, they are still about one order in magnitude above the optimal $\mathrm{MMD}^2$ value. The models with $L=5$ and $6$ layers are the first with sufficient expressivity to converge to the optimal value after approximately 70 epochs. 

Conversely, all QCBMs with \frg converge within about 50 epochs, with small variations at early stages of training. In fact, the $L=0$ circuit shows that this simple dataset can be learned with high precision without any entanglement. This shows that the choice of binary code can heavily influence the performance of a QCBM. 

Finally, the monotone Gray code \fmg performs better than the random code \frd, but does not achieve the same results as \fsc or \frg. The aggregated synthetic data shown in the bottom panels show that it consistently avoids certain bitstrings, while expressing others too often. Therefore, the good performance of \frg originates not only from the Gray property, but also from the inherent symmetry and the hierarchy between bits that \fmg is lacking. 

For both \fsc and \frg, there are bits of high and of low significance. Furthermore, the bits are ordered such that neighboring bits have similar significance. This is a perfect fit for the linear entangling topology employed in these experiments. The symmetry, together with the lower entanglement requirements due to the Gray property explain the advantage of \frg compared to the other codes.

The \frg QCBMs in fig.\ \ref{fig:Symmetric_Blob} exhibit two desirable properties: They converge quickly and to the optimal value. While those are in principle two independent properties, a good model should ideally both converge quickly and to a low minimum. Irrespective of the binary code, all models initially start with a probability distribution $p_\theta$ that is close to the uniform distribution and thus with a similar initial loss value. Furthermore, their achievable values are (on average) all bound from below by the same sampling constraints. A simple measure for a combination of both convergence speed and achieved minimum is thus given by the geometric mean of the loss function value over all epochs
\begin{align}
    Q_{\mathrm{loss}} = \exp\left( \sum_{i} \log \left( \mathrm{loss}(q, p_{\theta^i}) \right) \right) ~, \label{eq:quality}
\end{align}
where $p_{\theta^i}$ is the model at epoch $i$. The geometric mean is chosen here instead of the arithmetic mean to put less emphasis on the first epochs with very large loss values, which dominate the arithmetic mean.
For each model, fig.~\ref{fig:Symmetric_Blob} shows these values for the $\mathrm{MMD}^2$ loss $Q_{\mathrm{MMD}^2}$ on the right of each convergence plot. The above findings are thereby condensed in terms of a single number: The models with \frd struggle to represent the data well, whereas those with \fsc require several layers to recreate it well and those with \fmg would likely require several more. All models with \frg achieve very low $Q_{\mathrm{MMD}^2}$ values and only a very small dependence on the number of layers $L$. In the following we will focus on $Q_{\mathrm{MMD}^2}$ instead of the whole $\mathrm{MMD}^2$ convergence lines to judge a model's performance. 

Note that this dataset shows precisely those traits that are beneficial for \frg: (i) it is narrow and connected, such that only few adjacent bitstrings/ data points are needed, (ii) it is centered around the middle, such that \fsc requires very specific parameters that create the required entanglement between all qubits whereas \frg does not, and (iii) it is mirror-symmetric which further benefits \frg. To investigate whether the performances in fig.\ \ref{fig:Symmetric_Blob} translate to more general datasets as well, these traits are removed step-by-step in the following.

\subsection{Multiple Gaussian Distribution} \label{sec:Multiple_Gaussians}

In a first step, two more Gaussian distributions are added to the dataset, resulting in a total of three Gaussians of width $\nu = 0.03$. Unlike before, all of them are now randomly distributed in the interval $[-1, 1]$. If data points lie outside the interval $[-1, 1]$, the $x$ axis is rescaled to fit this interval. While each individual Gaussian is still mirror-symmetric, the overall distribution is not. This creates a level-playing field on which we can now compare \fsc and \frg using the performance metric $Q_{\mathrm{MMD}^2}$ defined in eq.\ \eqref{eq:quality}. Fig.~\ref{fig:3_Blobs} shows the $Q_{\mathrm{MMD}^2}$ values achieved by QCBMs with the four different binary codes. In an additional step, the number of qubits used to represent the data is varied.

The QCBMs with \frd and \fmg do not reach as small $Q_{\mathrm{MMD}^2}$ values as those with \fsc or \frg. Moreover, their performance decreases with increasing qubit count. This indicates that they quickly run into the barren plateau problem. As discussed above, this can be attributed to the missing or inappropriate structure in the mapped target probability distributions $q^{\frdinv}(b)$ and $q^{\fmginv}(b)$. Interestingly, the models with $n=16$ qubits also get worse when the number of layers $L$ is increased. This is remarkable, since all parameters are initialized close to zero, such that the initial distribution should be close to the uniform distribution regardless of the depth. If the parameters of the additional layers were fixed close to zero, and not optimized, these additional layers would not contribute to the model, and the deeper models would thus recreate the shallow one. Intuitively, deeper models with optimized additional layers should therefore perform at least as well as shallow ones. We investigate this phenomenon in appendix \ref{sec:appendix:deep_RC_models} and demonstrate that the optimization of the additional layers is actively harmful when using a random code. Small changes in the additional parameters lead to significantly different synthetic distributions $p^{\frd}_\theta$ (due to the random code), and move the model to flat parts of the loss landscape that are harder to optimize. 

\begin{figure*}
    \centering
    \includegraphics[width=\linewidth]{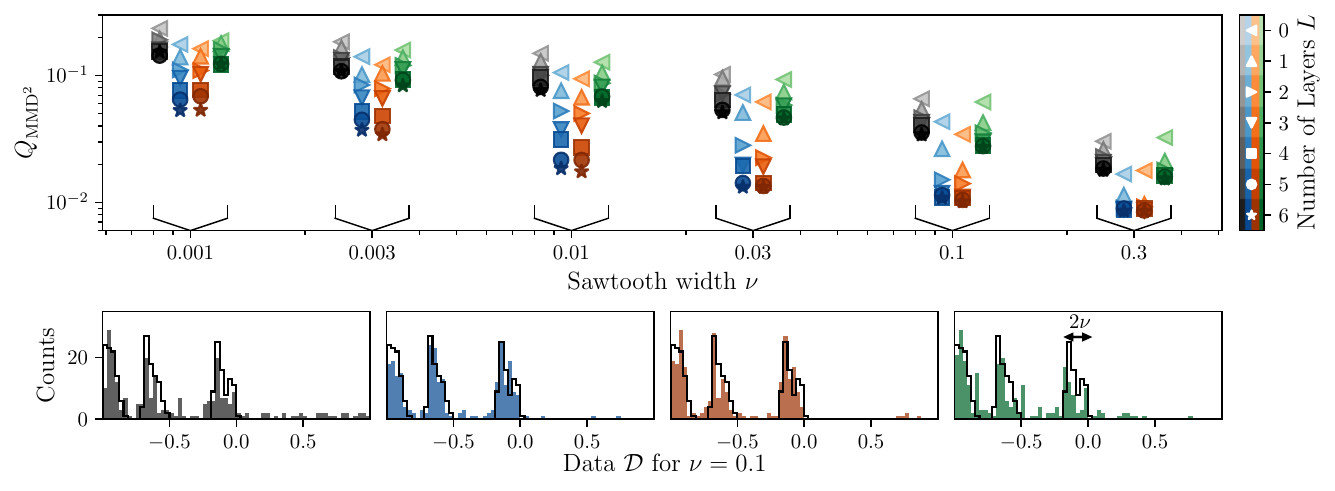}
    \caption{The simulation results for various QCBMs with 12 qubits, averaged over ten datasets drawn from different sawtooth distributions with randomly chosen means $\mu_i$. \textit{Top}: The $x$-axis scales the width $\nu$ of the sawtooths and the $y$-axis shows the $Q_{\mathrm{MMD}^2}$ scores achieved by the models. Models with binary codes \frd (black), \fsc (blue), \frg (orange), \fmg (green) are displayed from left to right. Like before, the markers and their lightness indicate the number of layers $L$. 
    \textit{Bottom}: 256 training (black line) and 256 synthetic data points (bars) for \textit{one} model with 12 qubits, $L=6$ layers and $\nu = 0.1$ trained for 100 epochs. For visibility, the data space $\mathcal{D}$ is discretized into $2^6$ (and not $2^{12}$) bins.  
    }
    \label{fig:3_Sawtooth}
\end{figure*}

\fsc and \frg show a clear dependence on the number of qubits $n$. For $n=6$ or $n=8$ qubits, they do not reach the same $Q_{\mathrm{MMD}^2}$ values as the models with more qubits. The discretization into $2^6$ representatives $x_j$, each representing an interval of $2 / 2^6 \approx 0.03$, is not enough to represent all details of the normal distributions with a standard deviation of $\nu = 0.03$. Similarly, it is larger than the smallest bandwidth parameter $\sigma = 0.003$
of the Gaussian kernels in the loss function
(see sec.~\ref{sec:loss_function}). All models with twelve or more qubits have a sufficient resolution and further increasing $n$ does not decrease the loss. At the same time, they also do not get worse with more qubits. Apparently, at least up to 16 qubits, barren plateaus can be avoided by combining a structure-preserving binary code with a loss function consisting of low-body and global terms. 

For almost all combinations of $n$ and $L$, the models with \frg outperform the corresponding models with \fsc. While this advantage is large for low circuit depths $L \leq 2$, it gets smaller for deeper circuits, where it is not as drastic as in fig.~\ref{fig:Symmetric_Blob}. This indicates that some of the advantage seen for the centered Gaussian dataset indeed originates in the central placement and the symmetry of the data both of which are disadvantageous for the \fsc. In total, \fsc achieves the lowest $Q_{\mathrm{MMD}^2}$ value in 6 of the 42 experiments, whereas  the \frg performs best in the remaining 36 experiments. 

The next section further changes the dataset to investigate whether the remaining advantage of \frg is due to the mirror-symmetry of each individual Gaussian distribution or the $2$-neighbor property of the \frg. Furthermore, we test whether \frg is flexible enough to fit non-continuous data as well.

\subsection{Multiple Sawtooth Distributions} \label{sec:Multiple_Sawtooths}

Following the results in sec.\ \ref{sec:Multiple_Gaussians}, we focus on models with $n=12$ qubits in this subsection. To remove the mirror-symmetry, three Gaussian distributions are replaced by three sawtooth distributions of width $2\nu$ and randomly chosen mean $\mu_i$. In this case, the target probability density function reads
\begin{align}
    q(x) = \sum_{i=1}^3\frac{\nu + \mu_i - x}{6 \nu^2} \Theta((\nu + \mu_i - x)(\nu - \mu_i + x)) \label{eq:q_sawtooth}
\end{align}
where $\Theta$ denotes the Heaviside function. Again, we sample 256 data points for the training. The width $2\nu$ is scaled from a very narrow $\nu=0.001$ distribution that covers only few representatives to $\nu=0.3$, in which a large part of the data space has some non-vanishing probability. The achieved $Q_{\mathrm{MMD}^2}$ scores of various models with the different binary codes and different numbers of layers $L$ are shown in fig.~\ref{fig:3_Sawtooth}. The datasets with large $\nu$ generally lead to smaller $Q_{\mathrm{MMD}^2}$ values. This is an effect of all models starting near a uniform distribution, which is closer to the target distribution for large $\nu$. The models therefore start from a smaller loss value and converge faster.

Like in the previous subsections, \frd performs worst for all dataset, followed by the \fmg. Out of the $42$ experiments, \fsc achieves the best $Q_{\mathrm{MMD}^2}$ value in $10$ experiments. Of those, five involve very narrow distributions $\nu = 0.001$. Since the resolution of all models is $2/2^{12} \approx 0.0005$, each sawtooth of width $2\nu=0.002$ then consists of four to five neighboring representatives, and the whole dataset of three such clusters. In this case, the Gray property in eq.\ \eqref{eq:Gray_Code_Property} is not as relevant. On the other side, also $3$ experiments with very broad distributions $\nu = 0.3$ are best with \fsc. In all experiments with intermediate width $\nu = 0.03 - 0.1$, \frg consistently outperforms almost all other binary codes. In total, \frg has the best $Q_{\mathrm{MMD}^2}$ values for $32$ out of the $42$ experiments. 

This shows that \frg provides an inductive bias that leads to a systematic improvement of QCBMs for a large range of datasets. At the same time, this example shows that, while the bias toward continuous data makes it easier to fit continuous data, it does not restrict the model to continuous functions. Changing the binary code therefore provides a straight-forward and easy-to-implement way of encoding an inductive bias into a generative model without limiting its expressivity.

\subsection{Robustness: Dimension, Loss and Ansatz} \label{sec:Higherdim_data}

So far, all simulations have used one-dimensional data, a single circuit ansatz, and the $\mathrm{MMD}^2$ loss. In the following, we test the robustness of the observed performance improvements by extending the distributions to multi-dimensional data. For the datasets, we draw $N=256$ data points from four randomly placed, two- and three-dimensional Gaussian distributions with with a standard deviation of $\nu = 0.03$. Every dimension is represented using $8$ (2D) and $5$ (3D) qubits, resulting in QCBMs with $n=16$ and $n=15$ qubits, respectively. In addition to the $\mathrm{MMD}^2$ loss, we repeat the simulations with the Sinkhorn divergence ($\mathrm{SHD}$, see appendix \ref{sec:appendix:SHD}) as loss function and evaluate these results using the metric defined in eq.~\eqref{eq:quality} for $\mathrm{SHD}$, $Q_\mathrm{SHD}$. Finally, we compare the HEA circuit with the all-to-all ansatz described in section \ref{sec:pqc}. Since deeper circuits significantly improved the results for these datasets, we include simulations with up to $L=9$ layers. 

\begin{figure}
    \centering
    \includegraphics[width=\linewidth]{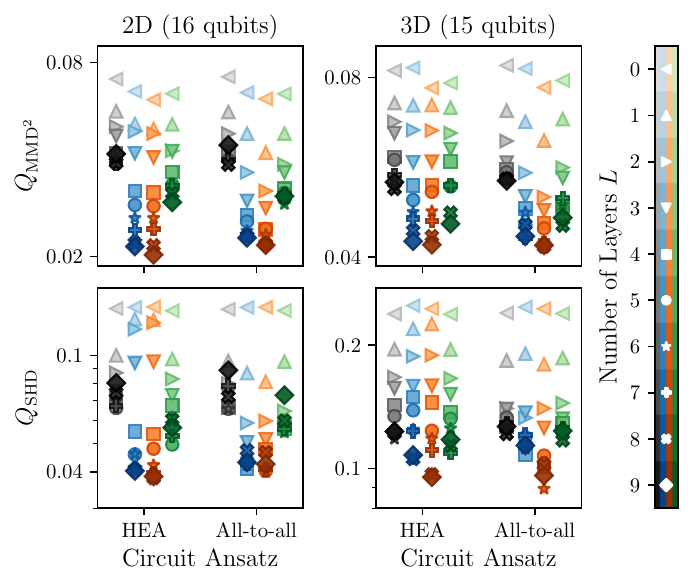}
    \caption{QCBMs trained on $d=2$ (left) and $d=3$ (right) dimensional data with the $\mathrm{MMD}^2$ (middle) and $\mathrm{SHD}$ (bottom) loss function. The training data consists of 256 data points drawn from four Gaussian distributions with standard deviation $\nu = 0.03$ that are randomly placed in $[-1, 1]^d$ with $d=2, 3$. The plots show the average $Q_{\mathrm{MMD}^2}$ and $Q_\mathrm{SHD}$ values of QCBMs trained on ten independently drawn datasets. The simulations use $8$ and $5$ qubits per dimension, resulting in $n=16$ and $n=15$ qubits for the two datasets. Two different circuit ansätze are used: the HEA and an all-to-all ansatz (see sec.~\ref{sec:pqc}). The number of layers is scaled between $L=0$ (no entanglement, light) and $L=9$ (dark). From left to right, the four groups of simulations use the random (black), standard (blue), reflected Gray (orange) and monotone Gray code (green). 
    Note that the y-axes of the different plots are not aligned and are not comparable due to the different setups. Consistent with the previous simulations, the figure indicates that the reflected Gray code typically yields the best QCBMs, independent of the loss function, number of dimensions and circuit ansätze, followed closely by the standard code.
    }
    \label{fig:2_3D_simulation}
\end{figure}

As shown in fig.~\ref{fig:2_3D_simulation}, the multi-dimensional results for HEA with $\mathrm{MMD}^2$ loss are similar to what we have observed in the one-dimensional case: QCBMs with \frd or \fmg perform significantly worse than those with \fsc or \frg. 
The synthetic data drawn from the 3D model with $L=9$ layers indicates that the higher-dimensional data might require even deeper circuits in order to capture the correlations between bits representing different dimensions. However, when comparing the different binary codes, the observations are consistent with the previous results. We see that the reflected Gray code typically achieves the best $Q_{\mathrm{MMD}^2}$ values, closely followed by the standard code. The advantage due to the inductive bias provided by \frg thus translates into the multi-dimensional setting. 

Compared to the $\mathrm{MMD}^2$, the $\mathrm{SHD}$ loss function leads to very similar results, with one minor exception in the low-depth regime and HEA circuit. For $L\leq 3$ and HEA, the random and monotone Gray code consistently perform \textit{better} than the standard or reflected Gray code. In this case, some of the models with \fsc and \frg are subject to a mode collapse, in which they recreate only a subset of the four Gaussians. The structure provided by these binary codes hinders them from randomly escaping the mode collapse, because that would require a correlated change of multiple qubits. While none of these low-depth models recreate the data well, the lack of structure allows models with random or monotone Gray code to produce data close to all Gaussians, which explains the qualitative difference. A similar effect can be observed for the $\mathrm{MMD}^2$ loss, but to a lesser extent. The models still exhibit mode collapse, but the $\mathrm{MMD}^2$ penalizes it less severely than the $\mathrm{SHD}$. As described in sec.~\ref{sec:loss_function} and appendix \ref{sec:appendix:SHD}, the $\mathrm{MMD}^2$ puts more emphasis on local similarities (especially for small bandwidth parameters). A model that reproduces a subset of the Gaussians accurately therefore still receives a relatively good score, whereas the $\mathrm{SHD}$, as a global transport-based measure, penalizes the missing modes more strongly. Once deeper circuits ($L \geq 4$) are used, the data is well-recreated by models with \fsc or \frg, whereas the models with \frd still produce many outliers. In this case, the models largely rank in the same order as they did in previous simulations, irrespective of the loss function. 

Lastly, we compare the different circuit ansätze and observe a trade-off between expressivity and trainability. When using few layers, the models with all-to-all connections typically reach lower $Q_{\mathrm{MMD}^2}$ or $Q_{\mathrm{SHD}}$ values than models with the HEA. This is expected, since the all-to-all layers distribute entanglement much faster than the nearest-neighbor layers of the HEA, and therefore fewer layers suffice to create correlations between all bits in the data. However, once deeper circuits are used, the HEA catches up, and in some cases even outperforms the all-to-all ansatz. The results with \frd and \fmg suggest that the models with all-to-all layers might run into the trainability issues discussed in appendix \ref{sec:appendix:deep_RC_models} with fewer layers than models with HEA. A plausible explanation is that the all-to-all layers create entanglement faster, such that the optimization moves the state away from the initial product state more rapidly and it reaches the flat regions in the optimization landscape with fewer layers. This leads to the observed differences between the circuit ansätze. Apart from the case with $\mathrm{SHD}$ and HEA ansatz discussed above, the reflected Gray code yields the best models in the majority of simulations for both circuit ansätze. 

These simulations are largely consistent with the previous results. The reflected Gray code leads to the best models in 54 out of 80 simulations, which is slightly lower than the $7/7$, $36/42$, and $32/42$ ratios found in the previous sections. This is due to the low-depth $\mathrm{SHD}$ simulations discussed above. The standard, monotone, and random codes show the best results in 10, 15, and 1 simulations, respectively, with the majority of the monotone Gray code's successes falling into the $L\leq 3$, $\mathrm{SHD}$ regime. A direct comparison between \fsc and \frg shows a score of 18 to 62. This confirms that the reflected Gray code is best suited for a large range of datasets, irrespective of the dimension, loss function, and circuit ansatz. 
The binary code acts as an intermediary between the circuit, which generates bitstrings, and the loss function, which acts on the data space. We therefore expect the same behavior for other loss functions and circuit ansätze, as long as the data is aligned with the inductive bias provided by \frg (and the circuit ansatz is not specifically designed to match a different binary code).


\section{Conclusion} \label{sec:conclusion}

The ability of QCBMs to learn and generalize from data fundamentally depends on structure that is present in the data. QCBMs and other generative QML models operate on a binary level, and for binary data, structure in the data directly translates to certain structures of the loss function and the model. 
The situation is different for other data types. Typically, real-world data is often structured and, for example, of numerical nature. The binary outputs $b$ of a generative QML model therefore have to be mapped to numerical data points $x_{f^{-1}(b)}$ by some binary code $f$.

In this manuscript, we have shown that the binary code is a crucial component of generative QML models that has previously been overlooked. We have investigated how it affects the training and the quality of the synthetic data created by quantum circuit Born machines with shallow circuits. Using a random code and a monotone Gray code, we have demonstrated that an ill-considered choice of the binary code can obscure the underlying structure of the data, hindering efficient and successful training. The standard code typically used for QCBMs avoids some of these obstacles by acting as an inductive bias toward translationally symmetric data, but it has shortcomings when it comes to data originating from continuous processes or exhibiting different symmetries. These shortcomings can be resolved with essentially no overhead by using the reflected Gray code, which outperforms the standard code in $80\%$ of our numerical simulations, especially in the case of the reflection-symmetric centered Gaussian distribution. This advantage persists even when removing global symmetries in the simulations with multiple Gaussians, choosing non-continuous sawtooth distributions with translational symmetries, or considering multi-dimensional continuous data. Additionally, we have seen that this is the case for both the loss functions and the two entangling topologies considered here. The reflected Gray code thus provides a simple and easy-to-implement way of improving the trainability of generative models on continuous distributions without restricting their expressivity, which renders it a promising candidate for many practically relevant scenarios. We stress, however, that different structures in the data might require different inductive biases and therefore different binary encodings. 


With this manuscript, we showed that the binary encoding used in generative QML is an important design choice just like the circuit ansatz or the entanglement patterns used in the model. Instead of always using the standard code, other codes should be considered as well. A well-chosen binary encoding can provide an inductive bias that speeds up and improves the learning of certain structures in the distributions, complementing other design choices such as the circuit ansatz or the entanglement connectivity.

\begin{acknowledgments}
This article was written as part of the Qu-Gov project, which was commissioned by the German Federal Ministry of Finance. It has been supported by the BMFTR Grant 13N17422 (snaQCs2025). AA and HK want to extend their gratitude to Kim Nguyen, Manfred Paeschke, Oliver Muth, Susanne Zimmermann and Yvonne Ripke for their continuous encouragement and support. The Bundesdruckerei GmbH and the Fraunhofer IAF have filed a joint patent in Germany (application number: 102025121128.0) with MK, TW, HK and AA listed as inventors. 
\end{acknowledgments}

\section*{Data availability}
Data generated during the current study are available from the corresponding author upon reasonable request. 

\section*{Author Contributions}
M.K. conceived the study, developed the methods, performed the analytical work and simulations, and wrote the manuscript. F.R., T.W., H.K., and A.A. initiated and coordinated the collaboration and defined the research area and application context. T.W. supervised and guided the study. H.K. and A.A. provided insight into synthetic data generation and contributed a motivating use case. All authors contributed through discussion and critical revision of the manuscript.

\appendix

\section{Sinkhorn Divergence} \label{sec:appendix:SHD}

The Sinkhorn divergence $\mathrm{SHD}^c_\epsilon$ is a loss function with regularization parameter $\epsilon$ and metric $c$ on the data space that can interpolate between the Wasserstein Distance ($\lim_{\epsilon \to 0}$) and the $\mathrm{MMD}^2$ with similarity measure $k(x, y) = -c(x,y)$ ($\lim_{\epsilon \to \infty}$). It can be written as \cite{feydy19,coyle20} 
\begin{align}
    \mathrm{SHD}^c_\epsilon(q, p_\theta) = W^c_\epsilon(q, p_\theta) - \frac{1}{2}W^c_\epsilon(q, q) - \frac{1}{2} W^c_\epsilon(p_\theta, p_\theta) ~, \label{eq:appendix:SHD}
\end{align}
where
\begin{align}
    W^c_\epsilon(q, p_\theta) = \mathbb{E}_{x\sim q}\left[ \beta(x) \right] + \mathbb{E}_{y\sim p_\theta}\left[ \gamma(y) \right] ~, \label{eq:appendix:regularized_Wasserstein_Sinkhorn_Potentials}
\end{align}
and $\beta$ and $\gamma$ are Sinkhorn potentials. For two datasets $\{x_i\}_{i=1, \dots, N}$ and $\{y_j\}_{j=1, \dots, M}$, the potentials evaluated at the data points $\beta_i = \beta(x_i)$ and $\gamma_j = \gamma(y_j)$ can be computed using the \textit{Sinkhorn algorithm}. Initially setting $\beta_i = 0 = \gamma_j$ for all $i$ and $j$, the potentials can be obtained by iteratively applying 
\begin{align}
    \beta_i &\leftarrow - \epsilon \log \left[\frac{1}{M} \sum_{j=1}^M \exp \left( \frac{\gamma_j - c(x_i, y_j)}{\epsilon} \right) \right] \label{eq:appendix:Sinkhorn_Algorithm_f} \\
    \gamma_j &\leftarrow - \epsilon \log \left[ \frac{1}{N} \sum_{i=1}^N \exp \left( \frac{\beta_i - c(x_i, y_j)}{\epsilon} \right) \right] ~, \label{eq:appendix:Sinkhorn_Algorithm_g}
\end{align}
which, for $\epsilon \geq 0.05$, typically converges after 10 -- 20 iterations \cite{feydy19}. 

For the symmetric problem $W^c_\epsilon(q, q)$ can be computed using a single update rule 
\begin{align}
    \beta_i &\leftarrow \frac{1}{2} \beta_i - \frac{\epsilon}{2} \log \left[ \frac{1}{N} \sum_{i^\prime} \exp \left( \frac{\beta_i - c(x_i, x_{i^\prime})}{\epsilon} \right) \right] \label{eq:appendix:Sinkhorn_algorithm_auto}
\end{align}
which, again for $\epsilon \geq 0.05$, typically converges after three iterations \cite{feydy19}.

Using eqs.~\eqref{eq:appendix:Sinkhorn_Algorithm_f}, \eqref{eq:appendix:Sinkhorn_Algorithm_g}, and \eqref{eq:appendix:Sinkhorn_algorithm_auto}, four sets of Sinkhorn potentials can be determined: $\beta_i$ and $\gamma_j$ for $W^c_\epsilon(q, p_\theta)$, as well as $\xi_i$ for $W^c_\epsilon(q,q)$ and $\tau_j$ for $W^c_\epsilon(p_\theta, p_\theta)$. The empirical $\mathrm{SHD}^c_\epsilon$ can then be assembled from equations \eqref{eq:appendix:SHD} and \eqref{eq:appendix:regularized_Wasserstein_Sinkhorn_Potentials}
\begin{align}
    \mathrm{SHD}^c_\epsilon(q, p_\theta) = \frac{1}{N} \sum_{i=1}^N \left( \beta_i - \xi_i \right) + \frac{1}{M} \sum_{j=1}^M \left( \gamma_j - \tau_j \right) ~. \label{eq:appendix:SHD_final_assembly_from_sinkhorn_potentials}
\end{align}

Using the parameter shift rule, the derivatives of $\mathrm{SHD}^c_\epsilon$ can be computed with
\begin{equation}
    \begin{aligned}
    \frac{\partial}{\partial \theta_l} \mathrm{SHD}^c_\epsilon(q, p_\theta) = \frac{1}{2} \Big(&\mathbb{E}_{x_k \sim p^f_{\theta^{+}_l}} \left[ \varphi(x_k) \right] \\
    - &\mathbb{E}_{x_k \sim p^f_{\theta^{-}_l}} \left[ \varphi(x_k) \right] \Big) ~,
    \end{aligned}
\end{equation}
where 
\begin{equation}
    \begin{aligned}
        \varphi(x_k) = &- \epsilon \log \left[ \frac{1}{N} \sum_{i=1}^N \exp \left( \frac{\beta_i - c(x_i, x_k)}{\epsilon} \right) \right] \\
        &+ \epsilon \log \left[ \frac{1}{M} \sum_{j=1}^M \exp \left( \frac{\tau_j - c(y_j, x_k)}{\epsilon} \right) \right] ~,
    \end{aligned} \label{eq:appendix:SHD_Derivative}
\end{equation}
which has been derived in references \cite{coyle20} and \cite{feydy19}.

Throughout the main text, we set $\epsilon = 0.05$ and $c(x,y) = ||x-y||_2$, and omit the subscript $\epsilon$ and superscript $c$ on $\mathrm{SHD}^c_\epsilon$ for simplicity. Note that $k(x,y) = -c(x,y)$ differs from eq.~\eqref{eq:Gaussian_Kernel}. In the limit $\epsilon \to \infty$, the $\mathrm{SHD}^c_\epsilon$ would therefore converge to a different $\mathrm{MMD}^2$ loss function. The difference between these loss functions would be that the one with $k(x,y)=-||x-y||_2$ decreases similarity between data points linearly with their distances, whereas the Gaussian kernel (especially with small bandwidth) puts more weight on small distances.

\section{Average Hamming distance between bitstrings of neighboring representatives} \label{sec:appendix:average_Hamming}
The average Hamming distance between neighboring integers $H(f(j), f(j+1))$ depends on the used binary code $f$. For the random code, $\frd(j)$ and $\frd(j+1)$ are unrelated and, apart from the condition $\frd(j) \neq \frd(j+1)$, completely random. On average, random codes thus have
\begin{align}
    \langle H(\frd(j), \frd(j+1)) \rangle_j = \frac{2^n}{2^n-1} \frac{n}{2} \approx \frac{n}{2}~.
\end{align}
For the standard code there are in total $2^n/2^k$ pairs of bitstrings with distance $H(\fsc(j), \fsc(j+1)) = k$, resulting in an average of
\begin{align}
    \langle H(\fsc(j), \fsc(j+1)) \rangle_j &= \frac{1}{2^n-1} \sum_{k=1}^n k \frac{2^n}{2^k} \\
    &= \frac{2^n}{2^n-1} \left( 2 - \frac{n+2}{2^n} \right) \\
    &\approx 2 ~,
\end{align}
where the second identity can be derived from the finite geometric series. For $z \neq 1$
\begin{align}
    \sum_{k=0}^n z^k = \frac{1 - z^{n+1}}{1-z} ~.
\end{align}
Taking the $z$-derivative of both sides and multiplying by $z$ yields
\begin{align}
    z\sum_{k=0}^n k z^{k-1} &= z\frac{- (n+1) z^n (1 - z) + 1 - z^{n+1}}{(1-z)^2} \\
    &= \frac{z - (n+1)z^{n+1} + n z^{n+2}}{(1-z)^2} ~. 
\end{align}
Inserting $z=\frac{1}{2}$ gives the identity used above. 

The next section shows that the reflected Gray code \frg has the $2$-neighbor property. By definition, all Gray codes $f_G$, including the reflected Gray \frg and monotone Gray code \fmg satisfy
\begin{align}
    \langle H(f_G(j), f_G(j+1)) \rangle_j = 1 ~. \label{eq:2n_property_BRGC}
\end{align}

\section{Proof that the reflected Gray code maps close integers to close bitstrings }\label{sec:appendix:2n_proof}

In this section, we show that \frg is a bijection and it satisfies eq.\ \eqref{eq:Gray_Code_Property}. For arbitrary $0 \leq j < 2^n$, \frg is defined via eq.\ \eqref{eq:rgc}. Bit $0\leq i < n$ is thus given by 
\begin{align}
    \frg(j)_i = \fsc(j)_i \oplus \fsc(j)_{i+1} ~, \label{appendix:eq:rgc_bitwise}
\end{align}
where, for simplicity, we add an $n+1$th bit $\fsc(j)_n = 0$. The standard code $\fsc(j)$ can be reconstructed iteratively from $\frg(j)$ by starting with $\fsc(j)_{n-1} = \frg(j)_{n-1}$ and applying 
\begin{align}
    \fsc(j)_{i} &= \fsc(j)_{i+1} \oplus \frg(j)_i \label{appendix:eq:sc}
\end{align}
for $i=n-2, \dots, 0$. The transformation between $\fsc$ and $\frg$ is therefore a one-to-one map and \frg directly inherits \fsc's bijectivity. 

It remains to be shown that \frg maps similar integers to similar bitstrings. When increasing an integer $j$ by $1$, the corresponding transformation of its binary representation $\fsc(j)$ by the standard code is to flip the least significant bit that has the value $\fsc(j)_i = 0$ with $0\leq i < n$ to $1$ and to flip all the bits $\fsc(j)_l$ with lower significancy $l =0, \dots, i-1$ from $1$ to $0$ 
\begin{align}
    \fsc(j) \oplus \fsc(j+1) = \fsc(2^{i+1}-1) ~. \label{appendix:eq:i+1}
\end{align}
The bitwise xor operation $\oplus$ commutes with the right shift $R[\cdot]$. Inserting eqs.\ \eqref{eq:rgc} and \eqref{appendix:eq:i+1} we find
\begin{align}
    &\frg(j) \oplus \frg(j+1) \\
    &= \fsc(j) \oplus R[\fsc(j)] \oplus \fsc(j+1) \oplus R[\fsc(j+1)] \notag \\
    &= \fsc(j) \oplus \fsc(j+1) \oplus R[\fsc(j) \oplus \fsc(j+1)] \\
    &= \fsc(2^{i+1}-1) \oplus \fsc(2^i-1) \\
    &= \fsc(2^i) ~,
\end{align}
which has a $1$ on bit $i$ and $0$s everywhere else. This shows that for arbitrary $j$, $\frg(j)$ and $\frg(j+1)$ differ in exactly one bit which concludes the proof.

\section{The convergence behavior of deep models with random code} \label{sec:appendix:deep_RC_models}

The models with $n=16$ qubits and random code in fig.~\ref{fig:3_Blobs} in section \ref{sec:Multiple_Gaussians} show an unexpected behavior: They get worse when their expressivity is increased beyond $L=3$ layers, and models with $L=6$ layers perform noticeably worse, on average. The figure also suggests that models with monotone Gray code show the same behavior once more qubits or layers are used. Only for the standard and reflected Gray codes, no such trend is visible for the chosen range of qubits and layers. 

\begin{figure}
    \centering
    \includegraphics[width=\linewidth]{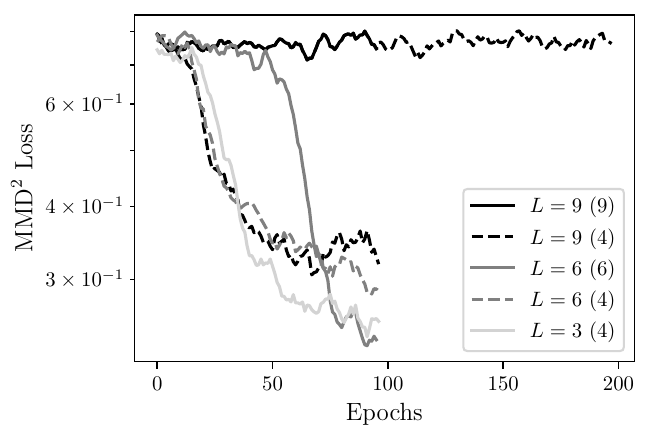}
    \caption{Training of different QCBMs with $\frd$ and $n=16$ qubits on one of the ten datasets drawn from three randomly placed, one-dimensional Gaussian distributions used in fig.~\ref{fig:3_Blobs}. For readability, the $y$-axis shows the rolling average of $5$ successive epochs. The different models all use the HEA circuit with different numbers of layers $L=3$ (light gray), $L=6$ (gray) or $L=9$ (black). The solid lines denote models that train all available $(L + 1) n$ parameters, whereas dashed lines denote models that were trained only on the last $4$ layers of the model (including the initial single-qubit rotation layer for $L=3$), and the other parameters are fixed to their initial value. The fully trained model with $L=9$ does not converge within $100$ epochs, which is why we added another $100$ epochs where only the last four layers are trained. }
    \label{fig:1D_Blobs_3D_half_training}
\end{figure}

In this section, we investigate this behavior. For this purpose, we conduct multiple simulations with one of the ten randomly drawn datasets with three Gaussian distributions from fig.~\ref{fig:3_Blobs}. To increase the effect, we also add a simulation with $L=9$ layers. The simulations are shown in fig.~\ref{fig:1D_Blobs_3D_half_training}. They largely confirm the observed trend. The models with $L=3$ converge fastest, those with $L=6$ converge significantly later, and those with $L=9$ do not find a suitable minimum in the loss function at all. 

For the models with $L=6$ and $L=9$, we additionally conducted simulations, in which only the parameters in the last $4$ single-qubit rotation layers are trained, and the others are fixed at their initial values, which were drawn uniformly between $-0.025$ and $0.025$. If these parameters were $0$, these simulations would exactly recreate the $L=3$ case, since the initial $|+\rangle$ state is invariant under CNOT gates. Instead, the small angles in the first layers will lead to a situation that is similar to the $L=3$ case, but with a slightly different initial state. Indeed, these models converge faster than their fully trained counterparts and behave similarly to the $L=3$ case, albeit converging to slightly worse loss function values. 

Lastly, we also used the $L=9$ model that was trained on all parameters for $100$ epochs and continued training the last $4$ layers for another $100$ epochs, while again fixing the initial layers. In this case, the training of only the last few layers does not converge.

The observations have a straightforward interpretation: the four models in which only four layers were trained differ only in their initial state. The $L=3$ model started in the $|+\rangle$ state and the $L=6$ model from a $|+\rangle$ state that was slightly perturbed by the initial layers with fixed small angles. Similarly, the $L=9~(4)$ model without pretraining started from a slightly more perturbed initial state. And the $L=9$ model that was pretrained on all layers likely started from a highly entangled, non-trivial state. The different performances therefore can be attributed to the difference in the initial states. The further the initial state moved away from the $|+\rangle$ state, the worse the model performed. This, once more, highlights the dependence of QML models on their initialization. Furthermore, it explains why the deeper models perform worse than shallow models. The optimization of the additional layers is actively harmful because it moves the model away from the initial state, where the later layers are easier to optimize, and toward flat areas in the loss landscape.

\bibliography{citations}

\end{document}